\documentclass[useAMS,usenatbib,usegraphicx]{mn2e}

\usepackage{times,epsfig}

\def\gsim{\mathrel{\raise0.35ex\hbox{$\scriptstyle >$}\kern-0.6em
\lower0.40ex\hbox{{$\scriptstyle \sim$}}}}
\def\lsim{\mathrel{\raise0.35ex\hbox{$\scriptstyle <$}\kern-0.6em
\lower0.40ex\hbox{{$\scriptstyle \sim$}}}}
\def\m@th{\mathsurround=0pt }
\def\eqalign#1{\null\,\vcenter{\openup1\jot \m@th
 \ialign{\strut\hfil$\displaystyle{##}$&$\displaystyle{{}##}$\hfil
 \crcr#1\crcr}}\,}
\def\micron{$\mu\textrm{m}$ }

\title[The physical scale of the far-IR in submillimetre galaxies]
        {The physical scale of the far-infrared emission in the most luminous submillimetre galaxies II: evidence for merger-driven star formation}
\author[Younger et al.]
             {Joshua D. Younger$^{\! 1,2}$, Giovanni G. Fazio$^{\! 3}$, Matthew  L. N. Ashby$^{\! 3}$, Francesca Civano$^{\! 3}$, \and 
             Martin Elvis$^{\! 3}$, Mark A. Gurwell$^{\! 3}$, Jia--Sheng Huang$^{\! 3}$, Daisuke Iono$^{\! 4}$, Alison B. Peck$^{\! 5}$, \and 
             Glen R. Petitpas$^{\! 3}$, Kimberly S. Scott$^{\! 6}$, David J. Wilner$^{\! 3}$, Grant. W. Wilson$^{\! 6}$, \& Min S. Yun$^{\! 6}$ \\
$^1$ Hubble Fellow \\
$^2$ School of Natural Sciences, Institute for Advanced Study, Einstein Drive, Princeton, NJ 08540, USA \\
$^3$ Harvard-Smithsonian Center for Astrophysics, 60 Garden Street, Cambridge, MA 02138, USA \\
$^4$ Nobeyama Radio Observatory, Minamimaki, Minamisaku, Nagano 384-1805, Japan \\
$^5$ Joint ALMA Office, El Golf 40, Las Condes, Santiago 7550108, Chile \\
$^6$ Astronomy Department, University of Massachusetts, Amherst, MA 01003}

\pagerange{000--000}

\begin{document}

\maketitle

\begin{abstract}

We present high-resolution 345 GHz interferometric observations of two extreme luminous ($L_{IR}\gsim 10^{13}$ $L_\odot$), submillimetre-selected galaxies (SMGs) in the COSMOS field with the Submillimeter Array (SMA).  Both targets were previously detected as unresolved point-sources by the SMA in its compact configuration, also at 345 GHz.  These new data, which provide a factor of $\gsim 3$ improvement in resolution, allow us to measure the physical scale of the far-infrared in the submillimetre directly.  The visibility functions of both targets show significant evidence for structure on $\sim 0.5-1$ arcsec scales, which at $z\gsim 1.5$ translates into a physical scale of $\sim5-8$ kpc.  Our results are consistent with the angular and physical scales of two comparably luminous objects with high-resolution SMA followup, as well as radio continuum and CO sizes.  These relatively compact sizes ($\lsim 5-10$ kpc) argue strongly for merger-driven starbursts, rather than extended gas-rich disks, as the preferred channel for forming SMGs.  For the most luminous objects, the derived sizes may also have important physical consequences; under a series of simplifying assumptions, we find that these two objects in particular are forming stars close to or at the Eddington limit for a starburst.

\end{abstract}


\section{Introduction}
\label{sec:intro}

In recent years, it has become increasingly clear that the most infrared(IR)-luminous galaxies pay an important role in the star formation history of the Universe.  Though they provide only a trivial contribution in the local Universe, both theory \citep{hopkins2009.ulirg,hopkins2010.sfh} and observations \citep{lefloch2005,perezgonzalez2005,magnelli2009,goto2010} have shown that by $z\gsim 1$ the total IR luminosity density is dominated by luminous ($10^{11} < L_{IR}/L_\odot < 10^{12}$; LIRGs), and at higher redshifts by ultralumious systems ($L_{IR} > 10^{12}$ $L_\odot$ ULIRGs).  Thus detailed study of these populations, and the processes driving their tremendous radiative output is critical to a thorough understanding of galaxy formation more generally.

Submillimetre-selected galaxies (SMGs), discovered in the first deep cosmological surveys at 850\micron \citep{smail1997,hughes1998,barger1998} by the Submillimetre Common User Bolometer Array \citep[SCUBA:][]{holland1999}, represent some of the most extreme objects in the high-redshift Universe \citep[for a review, see][]{blain2002}.  Owing to a strong negative $k$-correction, the selection function at $\approx 800-1000$\micron is flat from $z\sim 1-10$, and therefore provdes an unbiased view of star formation out to very high redshift \citep{blain1993}.  Though their bolometric energy output rivals that of luminous quasars, SMGs are several orders of magnitude more numerous at comparable redshifts \citep[$z\sim 2.5$;][]{chapman2005}.  At the same time, these two extreme populations are thought to be connected via an evolutionary sequence \citep[][]{hopkins2009.ulirg,narayanan2009.smg,narayanan2009.co,narayanan2009.dog} driving the formation of the most massive galaxies \citep{scott2002,blain2004,swinbank2006,swinbank2008,viero2009}.  Thus, in the context of a merger-driven cosmic cycle \citep[e.g.,][]{sanders1988a,hopkins2006,hopkins2007a,hopkins2007b}, SMGs represent the transition objects between star formation and AGN-dominated systems, and as such a compelling laboratory for testing models of galaxy formation and evolution in the most extreme environments.

\begin{table*}
\caption{Track Details}
\begin{center}
\begin{tabular}{cccccccccc}
 \hline
 \hline
Target & Configuration$^a$ & $u-v$ Coverage & Beam Size & Date & $<\tau_{\rm 225GHz}>$  & Obs. Time$^b$ & Reference$^c$ \\
 &  & [k$\lambda$] & [arcsec] & [dd.mm.yy] &  & [hrs] & \\
 \hline
AzTEC4 & COM & 20-75 & $2.71\times 2.12$ & 21.01.07 & 0.06 & 5.9 & Y07 \\
 & EXT & 50-250 & $0.86\times0.77$ & 23.02.09 & 0.05 & 3.9 & This work \\
\hline
AzTEC8 & COM & 20-75 & $2.69\times 2.19$ & 17.12.07 & 0.05 & 6.2 & Y09 \\
 & EXT & 50-250 & $0.86\times 0.55$ & 16.02.09,22.02.09 & 0.1,0.06 & 3.7,3.8 & This work \\
\hline
\hline
\end{tabular}
\end{center}
$^a$ For details on SMA configurations, see {\tt http://sma1.sma.hawaii.edu/status.html}.
$^b$ Total on--source integration time in that configuration.
$^c$ Y07: \citet{younger2007}; Y09: \citet{younger2009.aztecsma}
\label{tab:tracks}
\end{table*}

While their bolometric luminosity is thought to be primarily powered by star formation \citep{alexander2005,alexander2005b,alexander2008,valiante2007,menendez2007,menendez2009,pope2008b,momjian2010,serjeant2010}, SMGs could in principle represent one of two very different channels: a steady-state mode wherein large and extended \citep[e.g.,][]{kaviani2003,efstathiou2003} reservoirs of gas in disk galaxies fueled largely by cosmological gas accretion \citep[e.g.,][]{keres2005,keres2009,keres2009b,dave2009}\footnote{Here disk refers to a stable, gas-rich disk rather than one that is clump-unstable and highly turbulent \citep{elmegreen2008,dekel2009,cerevino2010}.  This scenario is discussed in \S~\ref{sec:multi} in the context of a potentially mult-component far-IR morphology.} yields a very high steady-state star formation rate, or interaction-driven events wherein close passages and/or mergers induce strong bars that centrally concentrate the gas supply \citep{barnes1996,hopkins2009.disksurvival}, leading to a brief and intense burst of nuclear star formation \citep[see also][]{hernquist1989a,barnes1991,mihos1994,mihos1996,cox2008,dimatteo2008}.  These two modes can be distinguished by the size of the star-forming region, with steady-state disks extending over tens of parsecs \citep[$R \approx 28 (S_\nu/{\rm 10\, mJy)^{1/2}}$ kpc at $z=2$:][]{kaviani2003} versus interaction-driven bursts which are much more concentrated. There is no {\it a-priori} reason to favor either scenario, and indeed both very gas-rich disks \citep[e.g.,][]{erb2006,daddi2009.fg,tacconi2010} and merger-driven, IR-luminous starbursts \citep[e.g.,][]{tacconi2008,walter2009} have been observed in the high-redshift Universe.

To date, observational estimates of the physical scale of star forming regions in SMGs prefer the interaction-driven scenario.   However, the majority of detections are in the radio, where by leveraging the remarkably tight far-IR radio correlation -- in which the total rest-frame 20cm luminosity provides a proxy for the total star formation rate among star-forming systems in the local Unvierse \citep{condon1992,yun2001,thompson2006,lacki2009a,lacki2009b} -- \citet{chapman2004}, \citet{biggs2008}, and \citet{momjian2010} inferred typical physical sizes of 5-10 kpc for SMGs.  At the same time, CO imaging \citep{neri2003,greve2005,tacconi2006,tacconi2008,daddi2008,daddi2009b,bothwell2010} tends to yield comparable sizes.  However, each of these techniques has its shortcomings: the spatially resolved far-IR/radio correlation is still not well understood \citep[see e.g.,][]{hippelein2003,murphy2006,tabatabaei2007}, and imaging redshifted CO emission in rotational lines with excitation requirements higher than the $J=1\rightarrow0$ transition may substantially underestimate the starburst scale \citep[][]{narayanan2008,ivison2010}.

\begin{figure}
\epsfig{figure=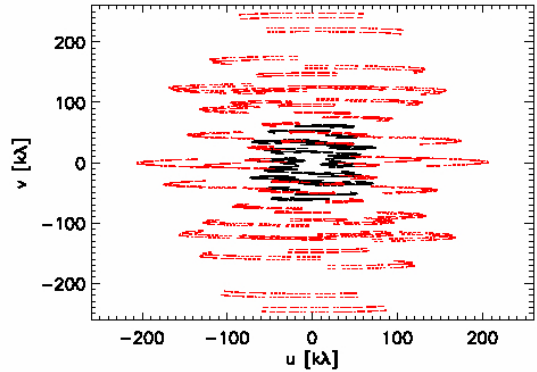,width=76mm}
\caption{The $u-v$ coverage for our high resolution interferometric imaging of AzTEC4 and AzTEC8 for both the compact (COM: black), and extended (EXT: red) configuration.  For further details, including weather conditions and on-source integration times, see Table~\ref{tab:tracks}.}
\label{fig:uv}
\end{figure}

Therefore, it is preferable to measure the starburst scale in the far-IR directly.  However, the beam sizes characteristic of single-dish instruments are often nearly an order of magnitude too large to probe the relevant spatial scales.  And, owing to technical constraints, even interferometric imaging is typically too coarse.  In fact, to date virtually all SMGs with interferometric followup are compact relative to the beam \citep[$\approx$ 2-3 arcsec FWHM;][]{dannerbauer2004,dannerbauer2008,iono2006,wang2007,younger2007,younger2008,younger2009.aztecsma,cowie2009}.  Recently, \citet{younger2008highres} imaged two of the brightest SMGs known with the Submillimetre Array \citep[SMA:][]{ho2004} in its most extended configurations -- the highest resolution submillimetre imaging of high-redshift starbursts achieved to date -- and found that their far-IR emission was extended on $\sim$few$\times$ kpc scales.  Such extremely bright objects were particularly interesting because at these size scales and luminosities the implied energetics start to run up against physical limits on the star formation rate imposed by the dynamical time of the gas \citep{elmegreen1999} or the effects of radiation pressure on the dust \citep{thompson2005,murray2005,murray2009,hopkins2010.maxsd}.  

In this work, we present similar observations of two other exceptionally bright, and therefore luminous objects in a 1.1mm blank-field survey of the COSMOS Field \citep{scott2008} performed with the AzTEC camera \citep{wilson2008}: AzTEC4 and AzTEC8.   In both cases, high-resolution follow-up observations provide a measurement (rather than an upper limit) on the scale of the far-IR emission, and therefore the starburst itself.  This paper is organized as follows: in \S~\ref{sec:obs} we summarize the observations and data reduction, in \S~\ref{sec:results} we present the results, in \S~\ref{sec:discuss} we discuss the implications, and in \S~\ref{sec:conclude} we conclude.  Throughout this work, we assume the 7-year WMAP cosmological model \citet{komatsu2010}, though any set of cosmological parameters within reason will not qualitatively change the interpretation of these results.

\begin{figure*}
\epsfig{figure=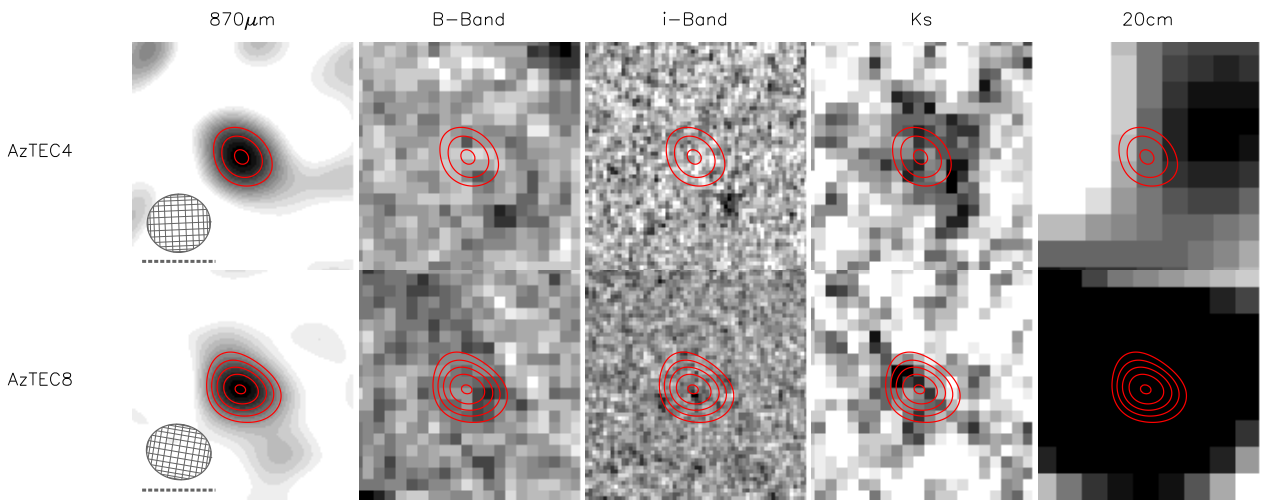,width=170mm}
\caption{Postage stamps for AzTEC4 (top) and AzTEC8 (bottom) including (left to right): 870\micron SMA (natural weighting), Subaru B-band, HST/ACS i-band, CFHT Ks-band, and VLA 20cm imaging.  For both sources, we only include SMA data from the EXT tracks.  The red contours indicate $3, 5,7,\ldots \times$ the r.m.s. noise level, the grey hashed ellipse the beam size, and the grey dashed line 1 arcsec for scale.}
\label{fig:stamps}
\end{figure*}

\section{Observations and data reduction}
\label{sec:obs}

The two targets -- AzTEC4 and AzTEC8 -- were selected from the SMA/AzTEC interferometric survey \citep{younger2007,younger2009.aztecsma} of bright sources in the AzTEC 1.1mm (270 GHz) survey of the COSMOS field \citep{scott2008}.   They were first identified in the AzTEC map with deboosted flux densities of $5.2^{+1.3}_{-1.4}$ and $5.5^{+1.3}_{-1.3}$ mJy, and were later detected by the SMA in its compact (COM) configuration at 870\micron (345 GHz) with flux densities of $14.4\pm 1.9$ and $19.7\pm 1.8$ mJy (assuming a point-source model).  We then targeted both sources with the SMA in its extended (EXT) configuration with a goal of measuring the source size.  The observing conditions for and an overview of all available data for the two targets are summarized in Table~\ref{tab:tracks}, and the $u-v$ coverage presented in Figure~\ref{fig:uv}.

A detailed description of the calibration strategy for the COM configuration tracks are provided in \citet{younger2007} and \citet{younger2009.aztecsma} for AzTEC4 and AzTEC8 respectively.  For the EXT tracks presented in this work, the receiver was tuned to 340 GHz in the LSB, and averaged with the USB for an effective bandwidth of 4 GHz centred at 345 GHz.  Passband calibration was performed using 3C84 \citep{bennett1962}, and primary flux calibration was done using either Ceres (on 16 and 23 Feb 2009) or Titan (on 22 Feb 2009).  The target was observed on a 10 minute cycle with two primary gain calibrators: 1058+015 ($\sim 2$ Jy; 15 degrees away) and 0854+201 ($\sim 6$ Jy; 24 degrees away).  Because Ceres is known to be variable at the $\sim 20-30\%$ level due to rotation \citep{altenhoff1994,redman1998,barrera2005}, we confirm this flux scale by checking that the flux density for 0854+201 derived using Ceres as the primary flux calibrator ($6.0\pm 0.3$ and $5.7\pm 0.3$ Jy) were consistent with those using Titan ($5.7\pm 0.3$ Jy).

In addition to the two primary targets, we observed a nearby test quasar -- J1008+063 ($\sim0.2$ Jy) which was 5 degrees away from the targets -- once every 60 minutes throughout the track to empirically verify the phase transfer and inferred source structure, and estimate the systematic positional uncertainty. This source is included in both the JVAS \citep{patnaik1992,browne1998} and VLBA Calibrator \citep{ma1998,beasley2002} surveys of compact, flat--spectrum radio sources, has an absolute position known to better than 20 mas, and was confirmed to be compact at 345 GHz by \citet{younger2008highres}.  

We also make use of extensive multiwavelength data in the COSMOS Field \citep[see][for an overview]{scoville2007}\footnote{More current information is available at {\tt http://cosmos.astro.caltech.edu/}.} including: Subaru ground based optical and near-IR \citep{taniguchi2007}, HST/ACS i--band \citep{koekemoer2007}, IRAC 3.6--8\micron and MIPS 24\micron \citep{sanders2007}, VLA 20cm \citep{schinnerer2007}, and {\it Chandra} X-ray \citep{elvis2009,puccetti2009} imaging.

\section{Results}
\label{sec:results}

\begin{figure*}
\epsfig{figure=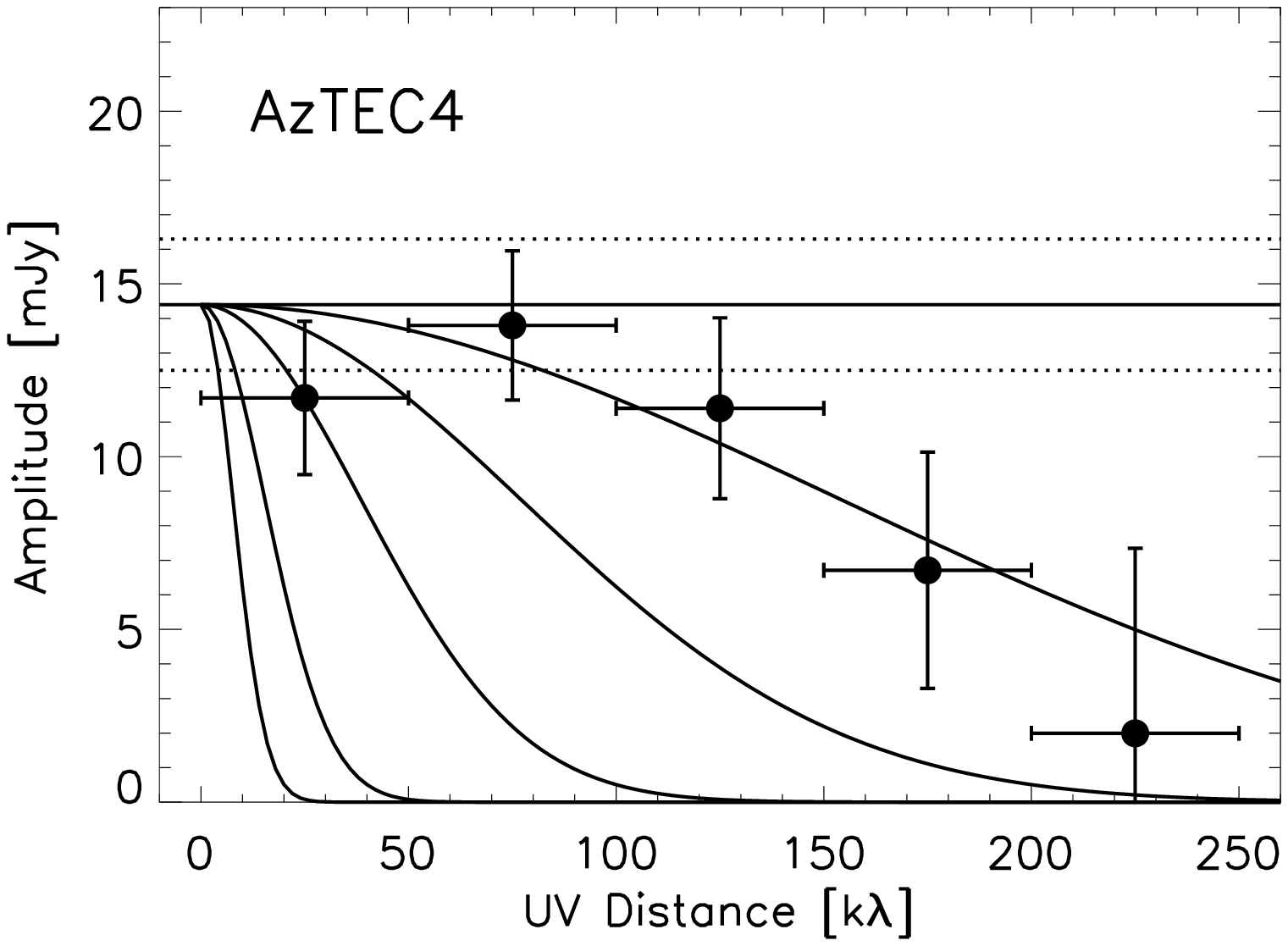,width=76mm}
\epsfig{figure=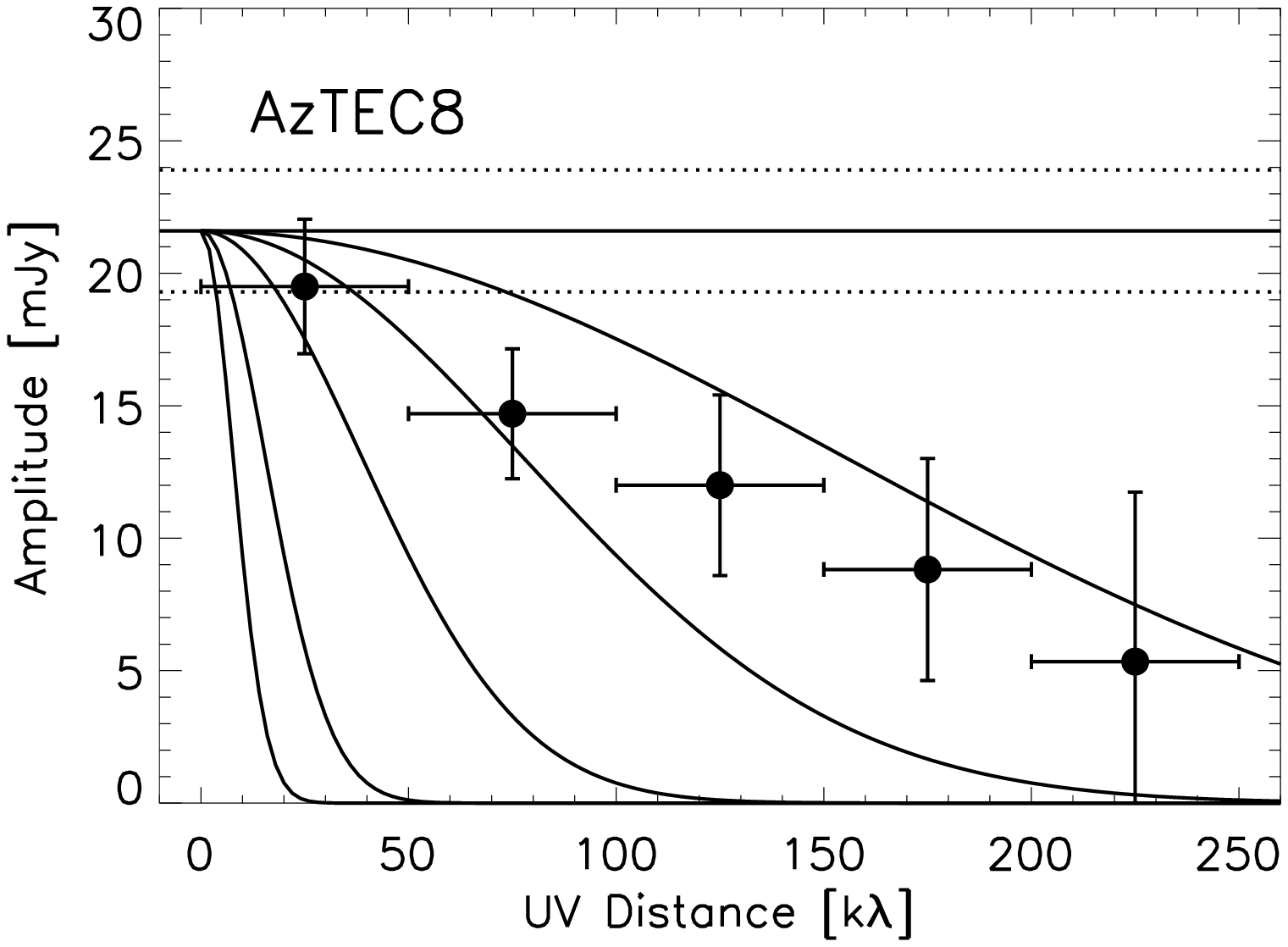,width=76mm} \\
\epsfig{figure= 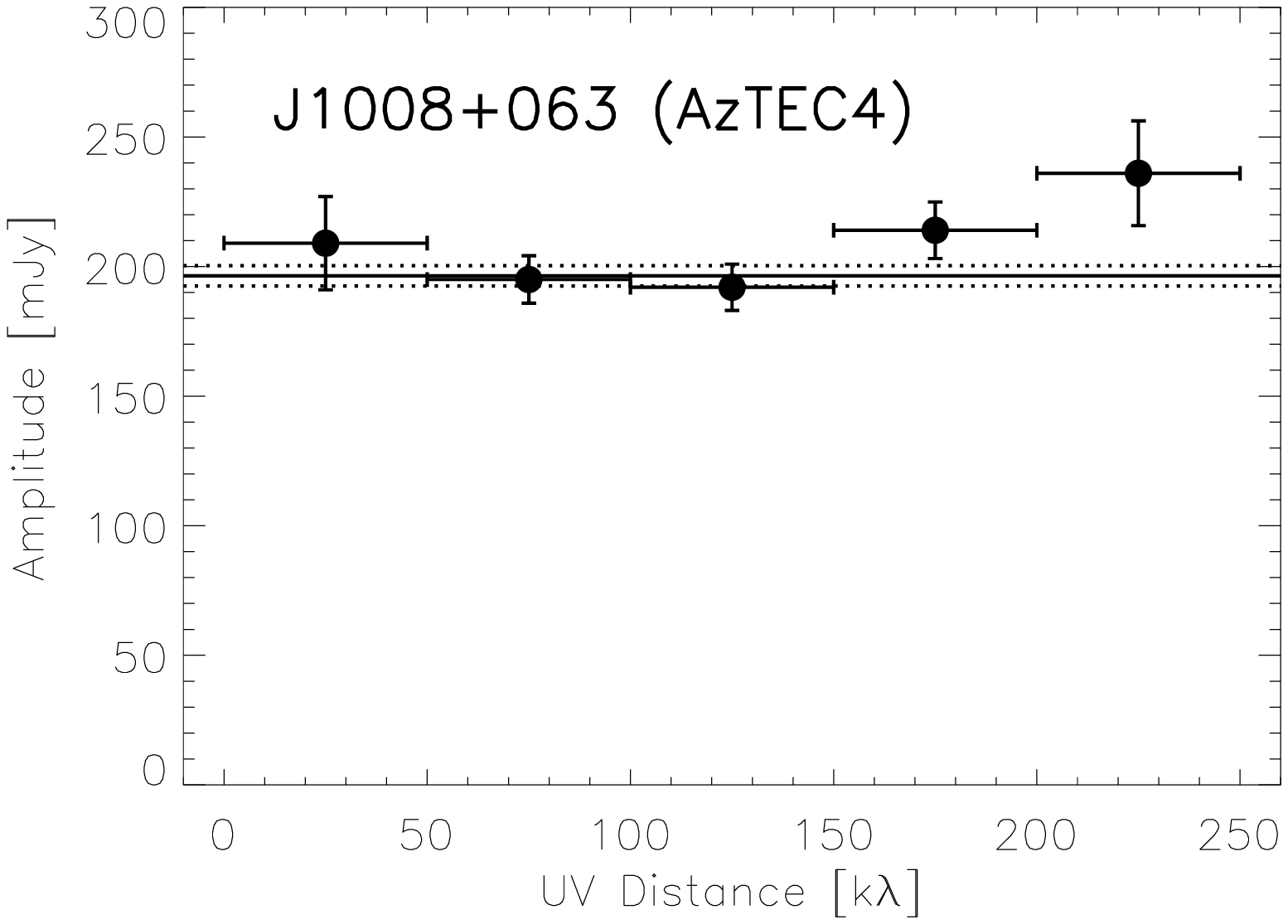,width=76mm}
\epsfig{figure= 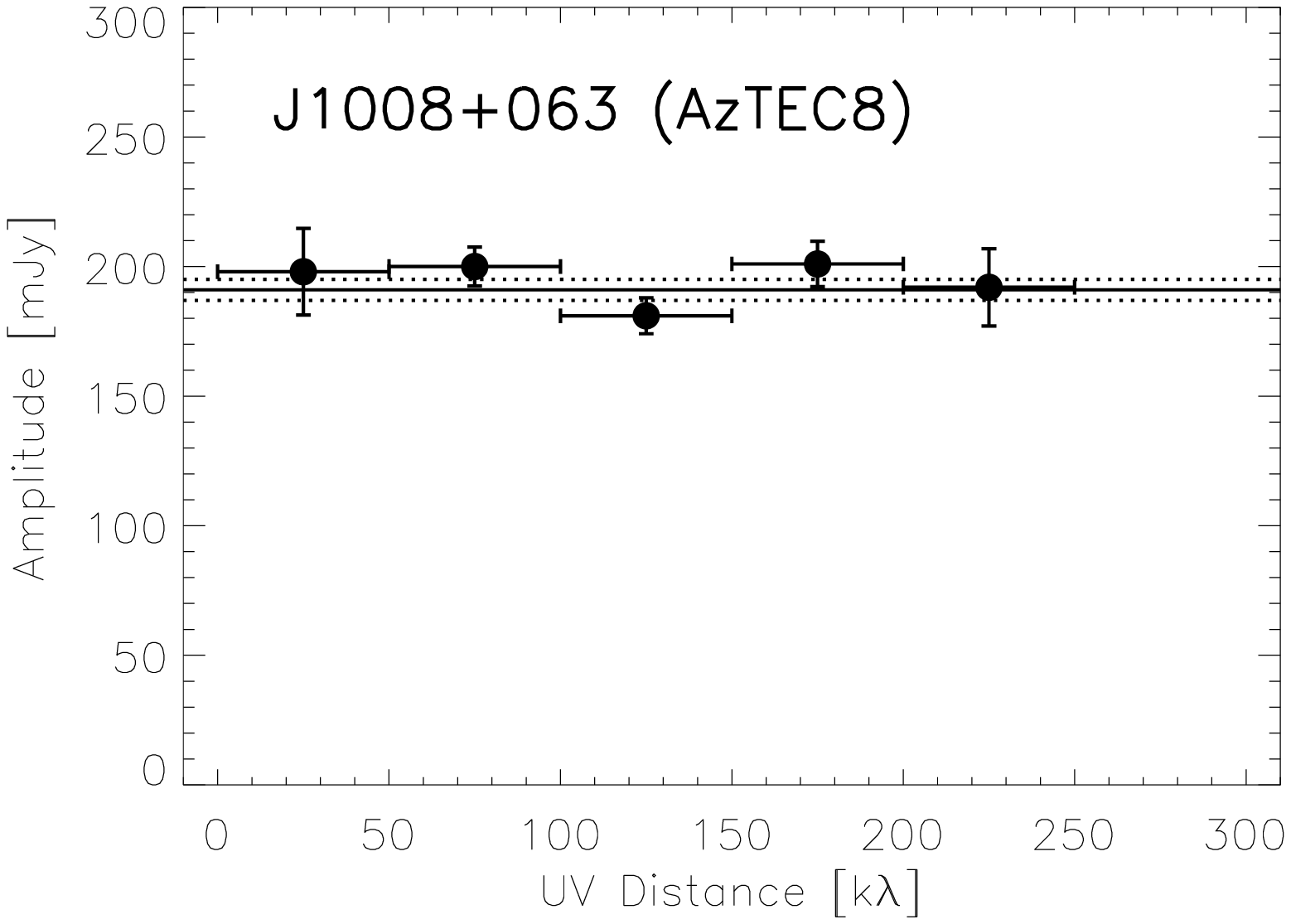,width=76mm}
\caption{Binned and vector-averaged real visibility amplitudes as a function of $u-v$ distance for the target (top row) and test quasar (bottom row).  The targets -- AzTEC4 (left) and AzTEC8 (right) -- and the calibration strategy are discussed in \S~\ref{sec:obs}.  For the targets we include all available data, including both COM and EXT tracks, details of which are provided in Table~\ref{tab:tracks} and the $u-v$ coverage presented in Figure~\ref{fig:uv}; for the test quasars we only include the EXT data.  For comparison, we provide a series of symmetric Gaussian source models with FWHM sizes of (left to right) 10, 5, 2, 1, 0.5 arcsec.  The flat solid and dotted lines indicate a point-source fit (including only the COM data for the targets, and only the EXT data for the test quasars) and the associated statistical uncertainty.  We find that both AzTEC4 and AzTEC8 show clear evidence of structure on $\approx 0.5-1$ arcsec scales, and the test quasars confirm the phase transfer.}
\label{fig:vis}
\end{figure*}

\begin{table*}
\caption{Positions and Source Structure}
\begin{center}
\begin{tabular}{cccccccccccc}
\hline
\hline
\tiny
Name & Config.$^a$ & Model & $\alpha$ & $\delta$ & $\Delta\alpha^b$ & $\Delta\delta^b$ & $F_{\rm 890\mu m}$ & $\theta^c_{\rm maj}$ & $\theta^c_{\rm min}$ & $\phi^d$  \\
 &  &  & [J2000] &[J2000] & [arcsec] & [arcsec] & [mJy] & [arcsec] & [arcsec] & [deg] \\
\hline
AzTEC4 & C & Point & 09:59:31.72 & +02:30:44.0 & 0.15 & 0.24 & $14.4\pm 1.9$ & \ldots & \ldots & \ldots \\
& C+E & Gaussian & 09:59:31.709 & +02:30:44.06 & 0.11 & 0.08 & $13.1\pm 1.8$ & $0.6\pm0.2$  & $0.4\pm0.2$ & 0 \\
& C+E & Disk & 09:59:31.709 & +02:30:44.06 & 0.09 & 0.07 & $13.1\pm 1.7$ & $1.0\pm0.4$  & $0.7\pm0.6$ & 30 \\
\hline
AzTEC8 & C & Point & 09:59:59.34 & +02:34:41.0 & 0.10 & 0.10 & $21.6\pm 2.3$ & \ldots & \ldots & \ldots \\
& C+E & Gaussian & 09:59:59.334 & +02:34:41.12 & 0.064 & 0.058 & $17.7\pm2.3$ & $0.6\pm0.2$  & $0.5\pm0.3$ & 40 \\
& C+E & Disk & 09:59:59.334 & +02:34:41.09 & 0.054 & 0.056 & $17.2\pm2.1$ & $1.0\pm0.5$  & $0.4\pm0.8$ & 20 \\
\hline
\end{tabular}
\end{center}
$^a$ See Table~\ref{tab:tracks} and Figure~\ref{fig:uv} for details.
$^b$ Combined statistical and systematic uncertainty, where the systematic uncertainty is estimated from the position of the test quasar.
$^c$ $\theta_{\rm maj}$ and $\theta_{\rm min}$ represent the FWHM or diameter of the major and minor axes for the Gaussian and elliptical disk models respectively.
$^d$ Position angle.
\label{tab:results}
\end{table*}

Imaging and model fitting for the calibrated visibility data was performed using the {\sc miriad} software package \citep{sault1995}.  We utilize a natural weighting scheme, yielding dirty maps for the EXT tracks that had r.m.s. noise levels of 1.6 and 1.8 mJy/beam for AzTEC4 and AzTEC8 respectively.  The peak flux density in each map was spatially consistent with detections in earlier COM tracks and highly statistically significant: for AzTEC4 the peak was 8.0 mJy/beam (S/N$\sim$5), and for AzTEC8 it was 12.8 mJy/beam (S/N$\sim$7).

Postage stamps of these maps, along with contours overlaid on optical, near-IR, and radio data are shown in Figure~\ref{fig:stamps}.  Neither source has a statistically significant detection in the optical or near-IR, though there is a marginal $\approx 23.3$ mag K-band detection associated with AzTEC4.  In the radio, as expected there is no emission coincident with AzTEC4 \citep[as in][]{younger2007} and the centroid of AzTEC8 is consistent with that of an $89\pm11$ $\mu$Jy source -- as in the COM data \citep[which has a comparable noise level;][]{younger2009.aztecsma} the nearby secondary radio counterpart of AzTEC8 is not detected in the EXT image.

We then combined the new EXT with existing COM data and inspected the visibility function for each target (see the top panel of Figure~\ref{fig:vis}).  When binned by $u-v$ distance and vector-averaged, both AzTEC4 and AzTEC8 show significance evidence for structure at $\gsim 100$ k$\lambda$, which corresponds to angular scales of $\lsim 0.6$ arcsec.  This is consistent with the images, which exhibits lower peaks in the higher resolution maps: for AzTEC4 the peak flux density in COM $13.2\pm 1.7$ mJy/beam versus $8.0\pm 1.8$ mJy/beam in EXT, and for AzTEC8 the peaks are $19.7\pm 1.8$ mJy/beam in COM and $12.8\pm 1.8$ mJy/beam in EXT.  In both cases, the test quasar empirically confirms that this structure is real, and not the result of decorrelation on the longest baselines (see bottom panels of Figure~\ref{fig:vis}). 

We then fit the calibrated visibility data to one of two simple models -- a 2-dimensional Gaussian or an elliptical disk -- the results of which are presented in Table~\ref{tab:results}.  The best-fit parameters for both sources show structure on $\approx 0.5-1$ arcsec scales at $\gsim 2\sigma$ confidence.  We also confirm that this size measurement is not artificially imposed by differing flux scales between the COM and EXT data by fitting the same source model to each set of calibrated visibilities independently: when we assume e.g. an elliptical Gaussian we find total flux densities (for AzTEC4 and AzTEC8 respectively, and including an estimate of the systematic uncertainty in the flux scale\footnote{Estimated from measurements of the flux density of 0854+201 which are available online as part of the SMA Calibrator List at {\tt http://sma1.sma.hawaii.edu/callist/callist.html}}) of $14.1\pm 2.5_{\rm stat}\pm 0.8_{\rm sys}$ and $17.9\pm 3.7_{\rm stat}\pm 0.9_{\rm sys}$ mJy for the COM data, and $9.2\pm 3.1_{\rm stat} \pm 0.5_{\rm sys}$ and $21.0\pm 4.8_{\rm stat} \pm 1.0_{\rm sys}$ mJy for the EXT data; that these are consistent between configurations to within the statistical and systematic uncertainties indicates that the size measurements listed in Table~\ref{tab:results} are robust.

\section{Discussion}
\label{sec:discuss}

\subsection{The physical scale of the far-infrared}
\label{sec:scale}

These angular scales are qualitatively consistent with size measurements derived from previous high-resolution far-IR imaging \citep{younger2008highres}, as well as other wavelengths including radio continuum \citep{chapman2004,biggs2008,momjian2010} and CO \citep{neri2003,tacconi2006,tacconi2008} imaging.  In our view, imaging in the far-IR directly is the most robust of these methods because it directly probes the obscured starburst; though the galaxy-wide far-IR/radio correlation is thought to apply -- at least in an qualitative sense -- at redshifts typical of SMGs \citep{garrett2002,gruppioni2003,appleton2004,boyle2007,younger2008.egsulirgs,ivison2009.firradio,murphy2009}, the spatially resolved correlation is not particularly well understood \citep{hippelein2003,murphy2006,tabatabaei2007}, while CO imaging is largely restricted to excited states with higher critical densities than the bulk of the star-forming gas, and therefore may be biased towards smaller sizes \citep{narayanan2008,ivison2010}.  However, given the agreement between these three methods -- now including twice as many objects with sizes measured in the far-IR directly -- we have newfound confidence that the general conclusion that the most luminous SMGs have typical angular scales of $\approx 0.5-1$ arcsec is robust.

To translate this into a physical size requires knowledge of the angular diameter distance.  Fortunately, at $1.5 \lsim z \lsim 5$ and assuming the 7-year WMAP $\Lambda$CDM cosmological model \citep{komatsu2010}, this quantity has a very weak scaling with redshift.  In general, the physical size ($\ell$) is approximately:
\begin{equation}
\ell \approx 11 \left (\frac{\theta}{\rm arcsec} \right ) (1+z)^{0.20}\,\,{\rm kpc}.
\end{equation}
Therefore, existing size measurements for SMGs suggest a typical physical scale of $\approx 4-8$ kpc at $z\sim 2-5$.  While there is clear evidence for $\gsim 20$ kpc sizes among some SMGs \citep{ivison2010}, the data presented in this work strongly argue against pure steady-state star formation fed by cosmological gas accretion \citep{keres2005,keres2009,keres2009b,dave2009} as the primary mechanism for producing SMGs, and rather favors the merger-driven scenario \citep{narayanan2009.smg,narayanan2009.co} in which interaction-driven starbursts give rise to these extreme objects.  This in turn fits into a more general evolutionary picture in which SMGs are transition objects \citep{hopkins2009.ulirg} between isolated, gas-rich spirals and luminous quasars, and the progenitors of passive, elliptical galaxies \citep[e.g.,][]{sanders1988a,hopkins2006,hopkins2007a,hopkins2007b}.  Furthermore, since these sizes are considerably larger than local ULIRGs \citep[see review by][]{sanders1996}, which themselves are powered primarily by major mergers \citep[e.g.,][]{scoville2000,kim2002,veilleux2002}, they are consistent with the canonical picture of SMGs as `scaled up' ULIRGs.

\subsection{Are the most luminous submillimetre galaxies Eddington-limited starbursts?} 

Size measurements are particularly interesting for the most luminous SMGs because their star formation is sufficiently rapid to run up against fundamental physical limitations.  Owing to a strong negative $k$-correction in the submillimetre \citep{blain1993,blain2002}, we can infer the total infrared luminosity in an approximate sense by assuming an Arp 220 template, which for $z\gsim 1$ yields $L_{FIR}/10^{12}\, L_\odot \approx 2 \times [S_{\rm 870\mu m}/{\rm mJy}]$ \citep[see also][]{neri2003,younger2008}.  By further assuming a \citet{salpeter1955} IMF\footnote{As noted in \citet{younger2008highres}, The inferred luminosity is known to be uncertain by a factor of $\sim 2 - 3$ due to variations in the dust temperature and emissivity \citet{blain2003}.  For example, if we assume a Mrk 231 template with significant AGN contribution to the farÐIR, the inferred farÐIR luminosity and SFR will be $\approx 40 - 60$\% lower \citep[e.g.,][]{stevens2005,huang2007}. Furthermore, using a \citet{kroupa2001} or \citet{chabrier2003} IMF will tend to lower any SFR by $\approx 40\%$ \citep{kennicutt1998,bell2003,bell2005}.}, this can be converted into a star formation rate ${\rm SFR} \approx 340 \times [S_{\rm 870\mu m}/{\rm mJy}] \, M_\odot \, {\rm yr}^{-1}$.  At the same time, Eddington-type arguments \citep{murray2005} -- under a series of simplifying assumptions such as a spherically symmetric geometry, an isothermal sphere density structure, a small volume filling factor for molecular gas, and again a \citet{salpeter1955} IMF -- yield a maximum star formation rate of:
\begin{equation}
{\rm SFR_{max}} = 900 \sigma_{400}^2 D_{\rm kpc} \kappa_{100}^{-1} \, \, M_\odot \, {\rm yr}^{-1}
\end{equation}
where $\sigma$ is the line-of-sight gas velocity dispersion in units of 400 km s$^{-1}$, $\kappa_{100}$ is the opacity in units of cm$^2$ g$^{-1}$ \citep[usually taken to be $\approx 1$;][]{murray2005,thompson2005}, and $D_{\rm kpc}$ is the characteristic physical scale of the starburst -- either the Gaussian FWHM or the elliptical disk diameter. 

\citet{younger2008highres} found that GN20 and AzTEC1 -- two of the most luminous SMGs known -- were potentially at or close to this Eddington limit.  In this work, we measured sizes for two comparably luminous objects: taking the flux densities inferred from the EXT+COM data, the ${\rm SFR}\approx 4400$ $M_\odot$ yr$^{-1}$ and $\approx 6000$ $M_\odot$ yr$^{-1}$ for AzTEC4 and AzTEC8 respectively.  Both have characteristic sizes of $\ell\approx 5-8$ kpc -- where the range represents uncertainty arising from a Gaussian versus disk source structure -- yielding a maximal star formation rate of ${\rm SFR_{max}}\approx 3600-7200$ $M_\odot$ yr$^{-1}$. Therefore, both objects appear to be forming stars at or near the Eddington limit for a starburst.


There are, however, a number of important caveats to consider:
\begin{itemize}
\item[1.] {\it Uncertainties in estimating the star formation rate}: There are few observational constraints on the stellar IMF at high-redshift and in extreme environments \citep[e.g.,][]{fardal2007,dave2008,VanDokkum2008,baugh2005,swinbank2008,tacconi2008} and the shape of the far-IR SED as a function of luminosity \citep[e.g.,][]{dale2001,dale2002,chary2001,lagache2003}.
\item[2.] {\it The volume filling factor of molecular gas}: In the optically thick limit -- in which the molecular gas has a volume filling factor near unity -- the Eddington limit is an order of magnitude higher.  A number of recent studies have found that this limit describes even local ULIRGs \citep{scoville1991,downes1993,downes1998,solomon1997} and therefore we might expect it to apply well to the much more extreme environments in SMGs.   This does, however, require a rather low effective dust temperature; in the optically thick limit, the brightness temperature at 345 GHz ($T_B$) is equal to the physical temperature ($T_d$).  While the sizes and flux densities of AzTEC4 and AzTEC8 require $T_d \approx 5 (1+z)$ {\sc k} \citep[see Figure~5 and discussion in \S~4 of][]{younger2008highres}, which at $z\sim 2-4$ is rather low compared to other SMGs for which $T_d$ has been measured independently via far-IR SED fitting \citep[$<T_d> \sim 35$ {\sc k} for typical SMGs, and the most luminous objects may be preferentially hotter:][]{kovacs2006,coppin2008.sharcii}, it is certainly not out of the question.  
\item[3.] {\it The depth of the potential well}: The maximal star formation rate has a strong dependence on the gravitational potential, which enters as $\sigma_{400}^2$.  Dynamical masses measured via CO spectroscopy have found that, in the mean $\sigma_{400} \approx 1$ for SMGs \citep{greve2005,tacconi2006}.  However, merger-driven models for SMGs suggest that the brightest 850\micron sources are also the most massive \citep{narayanan2009.smg}, and therefore would likely have $\sigma_{400}\gsim 1.5$ \citep[for GN20, $\sigma_{400} \approx 1.4$:][]{carilli2010} or so which could lead to significantly higher Eddington limits, even in the optically thin limit.
\end{itemize}
Generally speaking, however, we find that AzTEC4 and AzTEC8 are extended on physical scales comparable to the other two SMGs presented in \citet{younger2008highres}, and are potentially radiating at or close to the Eddington limit for a starburst.

\subsection{Obscured AGN versus star formation}
\label{sec:multi}

Though in general SMGs are thought to be star formation dominated  \citep{alexander2005,alexander2005b,alexander2008,valiante2007,menendez2007,menendez2009,pope2008b,momjian2010,serjeant2010}, in principle they could contain a significant contribution from an obscured AGN.  In fact, a preliminary analysis of recent very long baseline interferometry observations provides strong evidence that the radio continuum in at least some SMGs is powered primarily by an ultracompact AGN core \citep{biggs2009.evn,younger2010.evnproc}.  Clearly if this was the case for either AzTEC4 or AzTEC8 it would severely compromise their interpretation as extreme starbursts at or near their Eddington limit -- the Eddington limit for a $\sim 10^9$ $M_\odot$ supermassive black is well in excess of $10^{13}$ $L_\odot$.  

\begin{table}
\caption{Positions and Source Structure}
\begin{center}
\begin{tabular}{cccccccccccc}
\hline
\hline
Name & $F_{\rm 1, 890\mu m}^a$ & $F_{\rm 2, 890\mu m}^b$ & $\Delta\theta^c$ \\
& [mJy] & [mJy] & [arcsec] \\
\hline
AzTEC4 & $5.8\pm2.5$ & $6.8\pm2.6$ & $0.48\pm0.12$ \\
AzTEC8 & $13.0\pm1.6$ & $5.4\pm 1.6$ & $0.75\pm0.19$ \\
\hline
\hline
\end{tabular}
\end{center}
$^a$ Flux of the first component derived from fitting a two--component point--source model to the calibration visibilities.
$^b$ Flux of the second component derived from fitting a two--component point--source model to the calibration visibilities.
$^c$ The separation of the two fitted components.
\label{tab:two_points}
\end{table}

Recent X-ray imaging of the COSMOS field \citep[the C-COSMOS Survey:][]{elvis2009} provides some constraints on the AGN content of these objects; though starbursts also produce significant X-ray emission, a detection in the hard band ($2-8$ keV) would be strong evidence for the presence of a buried AGN -- particularly at high-redshift.   Therefore, we have examined the C-COSMOS X-ray imaging data for AzTEC4 and AzTEC8.  While AzTEC8 shows no evidence for a detection, AzTEC4 exhibits a tentative hard X-ray source.  A formal source extraction \citep[as in][]{puccetti2009} yields net counts in the hard band of $5.5\pm 2.7$ cts, which translates into a flux of $F_{HX} = 8.0\pm 3.9$ erg cm$^{-2}$ s$^{-1}$ -- a $\sim 2\sigma$ detection.  If we assume the bolometric corrections of \citet{hopkins2007.templateqso}, this implies an AGN with bolometric luminosity $L_{bol}/10^{11}\, L_\odot = 1.2\pm 0.5$, $3.4\pm 1.4$, and $7\pm2.4$ at $z=2$, 3, and 4 respectively, which translates into $M_{BH}\eta_{\rm edd} /10^7 \, M_\odot = 0.4\pm 0.2$, $1.0\pm 0.4$, and $2.1 \pm 0.8$ where $\eta_{edd}$ is the Eddington ratio of the AGN \citep[typically near unity during the peak of the starburst:][]{hopkins2005d,Hopkins2005c}.  While we cannot rule out significant X-ray absorption (Compton-thick in the case of non-detections), at these redshifts and energies the optical depths are unlikely to be extreme.  Therefore, the X-ray data suggests that AGN do not contribute significantly to the IR luminosity of AzTEC4 or AzTEC8.

\subsection{A multi-component source structure?}
\label{sec:multi}

The visibility functions for AzTEC4 and AzTEC8 show clear evidence for structure on $\approx 0.5-1$ arcsec scales (see Figure~\ref{fig:vis}), and fitting a model to the visibility data yields a statistically significant size measurement (see Table~\ref{tab:results}).  However, these observations are not sufficiently high-resolution, nor do they have sufficient signal-to-noise to distinguish an extended source structure from multiple compact components separated by less than the beam size.  When we fit a dual point-source model to the visibility data, both sources -- especially AzTEC8 -- yield statistically significant ($\gsim 2\sigma$) measurements for the implied sub-component flux densities and separations (see Table~\ref{tab:two_points}).  Therefore, while the visibility functions are consistent with an extended source structure, the data could also indicate a multi-component source structure produced by either more compact starbursts or Compton-thick AGN.  However, the sizes listed in Table~\ref{tab:results} can be thought of as upper limits on the physical scale of the starburst in these objects, and therefore still argue against an extended star-forming disk.

\citet{carilli2010} have argued -- using the specific case of GN20 \citep[see also][]{pope2006,younger2008highres}-- that such a multi-component source structure, particular one that appears circularly supported with significant extra-nuclear star formation is inconsistent with a major merger; rather, they they find that it is indicative of clumpy star formation owing to local instabilities in turbulent, high-redshift disks \citep{elmegreen2008}.  This is, however, not the case.  First, both hydrodynamical simulations \citep{cerevino2010} and analytic arguments \citep{dekel2009} indicate that this mode of star formation is steady-state, while the specific star formation rate (SSFR) of GN20, for example, is $\approx 10-30$ Gyr$^{-1}$ \citep{carilli2010}; a long duty cycle at this high a SSFR is unphysical, and furthermore is inconsistent with cold-mode accretion rates from cosmological simulations \citep{keres2009,keres2009b}.  Second, significant non-nuclear CO emission is not inconsistent with the merger scenario, particular one involving more than two participating galaxies \citep[though admittedly a more extreme case, see e.g. ][]{narayanan2006} .  Finally, should the clumps appear to exhibit disky kinematics, this would also be consistent with such a gas-rich merger; hydrodynamical simulations have shown that disky gas kinematics are generically conserved when the progenitors are extremely gas-rich \citep{robertson2004,hopkins2009.disksurvival}.

\section{Conclusion}
\label{sec:conclude}
 
We present high-resolution interferometric observations of two SMGs in the COSMOS field -- AzTEC4 and AzTEC8 -- performed with the SMA at 345 GHz.  The targets, two of the most luminous SMGs known, were previously detected as compact sources with the SMA at the same frequency in its COM configuration \citep[beam size $\approx 2.7\times 2.2$ arcsec:][]{younger2007,younger2009.aztecsma}.  These new observations, which offer a factor of $\gsim 3$ improvement in resolution, allow us to measure the angular size of the two targets.  The visibility functions show significant evidence of structure on angular scales of $\approx 0.5-1$ arcsec, in agreement with the sizes of two comparable objects measured by the SMA \citep{younger2008highres}, as well as radio continuum \citep{chapman2004,biggs2008,momjian2010} and CO \citep{neri2003,tacconi2006,tacconi2008,bothwell2010} imaging.  Owing to the weak scaling of angular diameter distance with redshift for $z\gsim 1.5$, we can convert this angular scale to physical units and find that the far-IR in these SMGs is extended over $\sim$few$\times$ kpc.  This provides evidence in favor of a merger-driven scenario for forming SMGs \citep{narayanan2009.co,narayanan2009.smg}, rather than extended gas-rich disks \citep{dave2009,ivison2010}.  For the most luminous objects, the derived sizes may also have important physical consequences; as with two comparable objects studied by \citet{younger2008highres}, under a series of simplifying assumptions the two targets in this study are forming stars at or near the Eddington limit for a starburst \citep{murray2005,thompson2005}.  
 
 \subsection*{Acknowledgements}
 
The Submillimeter Array is a joint project between the Smithsonian Astrophysical Observatory and the Academia Sinica Institute of Astronomy and Astrophysics and is funded by the Smithsonian Institution and the Academia Sinica.  This research has made use of data obtained from the Chandra Data Archive and software provided by the Chandra X-ray Center (CXC) in the application packages CIAO and Sherpa.  This research was based in part on data collected at Subaru Telescope, which is operated by the National Astronomical Observatory of Japan, as well as observations obtained with WIRCam, a joint project of CFHT, Taiwan, Korea, Canada, France, and the Canada-France-Hawaii Telescope (CFHT) which is operated by the National Research Council (NRC) of Canada, the Institute National des Sciences de l'Univers of the Centre National de la Recherche Scientifique of France, and the University of Hawaii.   We furthermore utilize observations made with the NASA/ESA Hubble Space Telescope.  The National Radio Astronomy Observatory is a facility of the National Science Foundation operated under cooperative agreement by Associated Universities, Inc, and the James Clerk Maxwell Telescope is operated by The Joint Astronomy Centre on behalf of the Science and Technology Facilities Council of the United Kingdom, the Netherlands Organisation for Scientific Research, and the National Research Council of Canada.   JDY acknowledges support from NASA through Hubble Fellowship grant \#HF-51266.01 awarded by the Space Telescope Science Institute,which is operated by the Association of Universities for Research in Astronomy, Inc., for NASA, under contract NAS 5-26555.  STScI is operated by the association of Universities for Research in Astronomy, Inc. under the NASA contract NAS 5-26555.  The HST COSMOS Treasury program was supported through NASA grant HST-GO-09822.  Work with the AzTEC data is supported, in part, by NSF grants AST 0828222 and AST 0907952.
 
\bibliographystyle{apj}
\bibliography{../smg}

\begin{thebibliography}{151}
\expandafter\ifx\csname natexlab\endcsname\relax\def\natexlab#1{#1}\fi

\bibitem[{{Alexander} {et~al.}(2005{\natexlab{a}}){Alexander}, {Smail},
  {Bauer}, {Chapman}, {Blain}, {Brandt}, \& {Ivison}}]{alexander2005b}
{Alexander}, D.~M., {Smail}, I., {Bauer}, F.~E., {Chapman}, S.~C., {Blain},
  A.~W., {Brandt}, W.~N., \& {Ivison}, R.~J. 2005{\natexlab{a}}, \nat, 434, 738

\bibitem[{{Alexander} {et~al.}(2005{\natexlab{b}})}]{alexander2005}
{Alexander}, D.~M. {et~al.} 2005{\natexlab{b}}, \apj, 632, 736

\bibitem[{{Alexander} {et~al.}(2008)}]{alexander2008}
---. 2008, \aj, 135, 1968

\bibitem[{{Altenhoff} {et~al.}(1994){Altenhoff}, {Johnston}, {Stumpff}, \&
  {Webster}}]{altenhoff1994}
{Altenhoff}, W.~J., {Johnston}, K.~J., {Stumpff}, P., \& {Webster}, W.~J. 1994,
  \aap, 287, 641

\bibitem[{{Appleton} {et~al.}(2004)}]{appleton2004}
{Appleton}, P.~N. {et~al.} 2004, \apjs, 154, 147

\bibitem[{{Barger} {et~al.}(1998)}]{barger1998}
{Barger}, A.~J. {et~al.} 1998, \nat, 394, 248

\bibitem[{{Barnes} \& {Hernquist}(1996)}]{barnes1996}
{Barnes}, J.~E. \& {Hernquist}, L. 1996, \apj, 471, 115

\bibitem[{{Barnes} \& {Hernquist}(1991)}]{barnes1991}
{Barnes}, J.~E. \& {Hernquist}, L.~E. 1991, \apjl, 370, L65

\bibitem[{{Barrera-Pineda} {et~al.}(2005){Barrera-Pineda}, {Lovell},
  {Schloerb}, \& {Carrasco}}]{barrera2005}
{Barrera-Pineda}, P.~S., {Lovell}, A.~J., {Schloerb}, F.~P., \& {Carrasco}, L.
  2005, in Revista Mexicana de Astronomia y Astrofisica Conference Series,
  Vol.~24, Revista Mexicana de Astronomia y Astrofisica Conference Series, ed.
  A.~M. {Hidalgo-G{\'a}mez}, J.~J. {Gonz{\'a}lez}, J.~M. {Rodr{\'{\i}}guez
  Espinosa}, \& S.~{Torres-Peimbert}, 188--191

\bibitem[{{Baugh} {et~al.}(2005){Baugh}, {Lacey}, {Frenk}, {Granato}, {Silva},
  {Bressan}, {Benson}, \& {Cole}}]{baugh2005}
{Baugh}, C.~M., {Lacey}, C.~G., {Frenk}, C.~S., {Granato}, G.~L., {Silva}, L.,
  {Bressan}, A., {Benson}, A.~J., \& {Cole}, S. 2005, \mnras, 356, 1191

\bibitem[{{Beasley} {et~al.}(2002){Beasley}, {Gordon}, {Peck}, {Petrov},
  {MacMillan}, {Fomalont}, \& {Ma}}]{beasley2002}
{Beasley}, A.~J., {Gordon}, D., {Peck}, A.~B., {Petrov}, L., {MacMillan},
  D.~S., {Fomalont}, E.~B., \& {Ma}, C. 2002, \apjs, 141, 13

\bibitem[{{Bell}(2003)}]{bell2003}
{Bell}, E.~F. 2003, \apj, 586, 794

\bibitem[{{Bell} {et~al.}(2005)}]{bell2005}
{Bell}, E.~F. {et~al.} 2005, \apj, 625, 23

\bibitem[{{Bennett}(1962)}]{bennett1962}
{Bennett}, A.~S. 1962, \memras, 68, 163

\bibitem[{{Biggs} {et~al.}(2009){Biggs}, {Younger}, \&
  {Ivison}}]{biggs2009.evn}
{Biggs}, A., {Younger}, J.~D., \& {Ivison}, R. 2009, in Proceedings of
  Panoramic Radio Astronomy: Wide-field 1-2 GHz research on galaxy evolution.
  June 2-5 2009. Groningen, the Netherlands. Published online at
  http://pos.sissa.it/cgi-bin/reader/conf.cgi?confid=89, id.20

\bibitem[{{Biggs} \& {Ivison}(2008)}]{biggs2008}
{Biggs}, A.~D. \& {Ivison}, R.~J. 2008, \mnras, 385, 893

\bibitem[{{Blain} {et~al.}(2003){Blain}, {Barnard}, \& {Chapman}}]{blain2003}
{Blain}, A.~W., {Barnard}, V.~E., \& {Chapman}, S.~C. 2003, \mnras, 338, 733

\bibitem[{{Blain} {et~al.}(2004){Blain}, {Chapman}, {Smail}, \&
  {Ivison}}]{blain2004}
{Blain}, A.~W., {Chapman}, S.~C., {Smail}, I., \& {Ivison}, R. 2004, \apj, 611,
  725

\bibitem[{{Blain} \& {Longair}(1993)}]{blain1993}
{Blain}, A.~W. \& {Longair}, M.~S. 1993, \mnras, 264, 509

\bibitem[{{Blain} {et~al.}(2002){Blain}, {Smail}, {Ivison}, {Kneib}, \&
  {Frayer}}]{blain2002}
{Blain}, A.~W., {Smail}, I., {Ivison}, R.~J., {Kneib}, J.-P., \& {Frayer},
  D.~T. 2002, \physrep, 369, 111

\bibitem[{{Bothwell} {et~al.}(2010)}]{bothwell2010}
{Bothwell}, M.~S. {et~al.} 2010, MNRAS, submitted [astro-ph/0912.1598]

\bibitem[{{Boyle} {et~al.}(2007)}]{boyle2007}
{Boyle}, B.~J. {et~al.} 2007, \mnras, 376, 1182

\bibitem[{{Browne} {et~al.}(1998){Browne}, {Wilkinson}, {Patnaik}, \&
  {Wrobel}}]{browne1998}
{Browne}, I.~W.~A., {Wilkinson}, P.~N., {Patnaik}, A.~R., \& {Wrobel}, J.~M.
  1998, \mnras, 293, 257

\bibitem[{{Carilli} {et~al.}(2010)}]{carilli2010}
{Carilli}, C.~L. {et~al.} 2010, ApJ, submitted [astro-ph/1002,3838]

\bibitem[{{Ceverino} {et~al.}(2010){Ceverino}, {Dekel}, \&
  {Bournaud}}]{cerevino2010}
{Ceverino}, D., {Dekel}, A., \& {Bournaud}, F. 2010, MNRAS, in press
  [astro-ph/0907.3271]

\bibitem[{{Chabrier}(2003)}]{chabrier2003}
{Chabrier}, G. 2003, \pasp, 115, 763

\bibitem[{{Chapman} {et~al.}(2005){Chapman}, {Blain}, {Smail}, \&
  {Ivison}}]{chapman2005}
{Chapman}, S.~C., {Blain}, A.~W., {Smail}, I., \& {Ivison}, R.~J. 2005, \apj,
  622, 772

\bibitem[{{Chapman} {et~al.}(2004){Chapman}, {Smail}, {Windhorst}, {Muxlow}, \&
  {Ivison}}]{chapman2004}
{Chapman}, S.~C., {Smail}, I., {Windhorst}, R., {Muxlow}, T., \& {Ivison},
  R.~J. 2004, \apj, 611, 732

\bibitem[{{Chary} \& {Elbaz}(2001)}]{chary2001}
{Chary}, R. \& {Elbaz}, D. 2001, \apj, 556, 562

\bibitem[{{Condon}(1992)}]{condon1992}
{Condon}, J.~J. 1992, \araa, 30, 575

\bibitem[{{Coppin} {et~al.}(2008)}]{coppin2008.sharcii}
{Coppin}, K. {et~al.} 2008, \mnras, 384, 1597

\bibitem[{{Cowie} {et~al.}(2009){Cowie}, {Barger}, {Wang}, \&
  {Williams}}]{cowie2009}
{Cowie}, L.~L., {Barger}, A.~J., {Wang}, W.-H., \& {Williams}, J.~P. 2009,
  \apjl, 697, L122

\bibitem[{{Cox} {et~al.}(2008){Cox}, {Jonsson}, {Somerville}, {Primack}, \&
  {Dekel}}]{cox2008}
{Cox}, T.~J., {Jonsson}, P., {Somerville}, R.~S., {Primack}, J.~R., \& {Dekel},
  A. 2008, \mnras, 384, 386

\bibitem[{{Daddi} {et~al.}(2009{\natexlab{a}})}]{daddi2009b}
{Daddi}, E. {et~al.} 2009{\natexlab{a}}, \apjl, 695, L176

\bibitem[{{Daddi} {et~al.}(2009{\natexlab{b}})}]{daddi2008}
---. 2009{\natexlab{b}}, \apj, 694, 1517

\bibitem[{{Daddi} {et~al.}(2009{\natexlab{c}})}]{daddi2009.fg}
---. 2009{\natexlab{c}}, ApJ, submitted [astro-ph/0911.2776]

\bibitem[{{Dale} \& {Helou}(2002)}]{dale2002}
{Dale}, D.~A. \& {Helou}, G. 2002, \apj, 576, 159

\bibitem[{{Dale} {et~al.}(2001){Dale}, {Helou}, {Contursi}, {Silbermann}, \&
  {Kolhatkar}}]{dale2001}
{Dale}, D.~A., {Helou}, G., {Contursi}, A., {Silbermann}, N.~A., \&
  {Kolhatkar}, S. 2001, \apj, 549, 215

\bibitem[{{Dannerbauer} {et~al.}(2008){Dannerbauer}, {Walter}, \&
  {Morrison}}]{dannerbauer2008}
{Dannerbauer}, H., {Walter}, F., \& {Morrison}, G. 2008, \apjl, 673, L127

\bibitem[{{Dannerbauer} {et~al.}(2004)}]{dannerbauer2004}
{Dannerbauer}, H. {et~al.} 2004, \apj, 606, 664

\bibitem[{{Dav{\'e}}(2008)}]{dave2008}
{Dav{\'e}}, R. 2008, \mnras, 385, 147

\bibitem[{{Dav{\'e}} {et~al.}(2009){Dav{\'e}}, {Finlator}, {Oppenheimer},
  {Fardal}, {Katz}, {Kere{\v s}}, \& {Weinberg}}]{dave2009}
{Dav{\'e}}, R., {Finlator}, K., {Oppenheimer}, B.~D., {Fardal}, M., {Katz}, N.,
  {Kere{\v s}}, D., \& {Weinberg}, D.~H. 2009, MNRAS, submitted
  [astro-ph/0909.4078]

\bibitem[{{Dekel} {et~al.}(2009){Dekel}, {Sari}, \& {Ceverino}}]{dekel2009}
{Dekel}, A., {Sari}, R., \& {Ceverino}, D. 2009, \apj, 703, 785

\bibitem[{{Di Matteo} {et~al.}(2008){Di Matteo}, {Bournaud}, {Martig},
  {Combes}, {Melchior}, \& {Semelin}}]{dimatteo2008}
{Di Matteo}, P., {Bournaud}, F., {Martig}, M., {Combes}, F., {Melchior}, A., \&
  {Semelin}, B. 2008, \aap, 492, 31

\bibitem[{{Downes} \& {Solomon}(1998)}]{downes1998}
{Downes}, D. \& {Solomon}, P.~M. 1998, \apj, 507, 615

\bibitem[{{Downes} {et~al.}(1993){Downes}, {Solomon}, \&
  {Radford}}]{downes1993}
{Downes}, D., {Solomon}, P.~M., \& {Radford}, S.~J.~E. 1993, \apjl, 414, L13

\bibitem[{{Efstathiou} \& {Rowan-Robinson}(2003)}]{efstathiou2003}
{Efstathiou}, A. \& {Rowan-Robinson}, M. 2003, \mnras, 343, 322

\bibitem[{{Elmegreen}(1999)}]{elmegreen1999}
{Elmegreen}, B.~G. 1999, \apj, 517, 103

\bibitem[{{Elmegreen} {et~al.}(2008){Elmegreen}, {Bournaud}, \&
  {Elmegreen}}]{elmegreen2008}
{Elmegreen}, B.~G., {Bournaud}, F., \& {Elmegreen}, D.~M. 2008, \apj, 688, 67

\bibitem[{{Elvis} {et~al.}(2009)}]{elvis2009}
{Elvis}, M. {et~al.} 2009, \apjs, 184, 158

\bibitem[{{Erb} {et~al.}(2006){Erb}, {Steidel}, {Shapley}, {Pettini}, {Reddy},
  \& {Adelberger}}]{erb2006}
{Erb}, D.~K., {Steidel}, C.~C., {Shapley}, A.~E., {Pettini}, M., {Reddy},
  N.~A., \& {Adelberger}, K.~L. 2006, \apj, 646, 107

\bibitem[{{Fardal} {et~al.}(2007){Fardal}, {Katz}, {Weinberg}, \&
  {Dav{\'e}}}]{fardal2007}
{Fardal}, M.~A., {Katz}, N., {Weinberg}, D.~H., \& {Dav{\'e}}, R. 2007, \mnras,
  379, 985

\bibitem[{{Garrett}(2002)}]{garrett2002}
{Garrett}, M.~A. 2002, \aap, 384, L19

\bibitem[{{Goto} {et~al.}(2010)}]{goto2010}
{Goto}, T. {et~al.} 2010, A\&A, in press [astro-ph/1001.0013]

\bibitem[{{Greve} {et~al.}(2005)}]{greve2005}
{Greve}, T.~R. {et~al.} 2005, \mnras, 359, 1165

\bibitem[{{Gruppioni} {et~al.}(2003)}]{gruppioni2003}
{Gruppioni}, C. {et~al.} 2003, \mnras, 341, L1

\bibitem[{{Hernquist}(1989)}]{hernquist1989a}
{Hernquist}, L. 1989, \nat, 340, 687

\bibitem[{{Hippelein} {et~al.}(2003)}]{hippelein2003}
{Hippelein}, H. {et~al.} 2003, \aap, 407, 137

\bibitem[{{Ho} {et~al.}(2004){Ho}, {Moran}, \& {Lo}}]{ho2004}
{Ho}, P.~T.~P., {Moran}, J.~M., \& {Lo}, K.~Y. 2004, \apjl, 616, L1

\bibitem[{{Holland} {et~al.}(1999)}]{holland1999}
{Holland}, W.~S. {et~al.} 1999, \mnras, 303, 659

\bibitem[{{Hopkins} {et~al.}(2008{\natexlab{a}}){Hopkins}, {Cox}, {Kere{\v s}},
  \& {Hernquist}}]{hopkins2007b}
{Hopkins}, P.~F., {Cox}, T.~J., {Kere{\v s}}, D., \& {Hernquist}, L.
  2008{\natexlab{a}}, \apjs, 175, 390

\bibitem[{{Hopkins} {et~al.}(2009){Hopkins}, {Cox}, {Younger}, \&
  {Hernquist}}]{hopkins2009.disksurvival}
{Hopkins}, P.~F., {Cox}, T.~J., {Younger}, J.~D., \& {Hernquist}, L. 2009,
  \apj, 691, 1168

\bibitem[{{Hopkins} \& {Hernquist}(2010)}]{hopkins2010.sfh}
{Hopkins}, P.~F. \& {Hernquist}, L. 2010, \mnras, 402, 985

\bibitem[{{Hopkins} {et~al.}(2005{\natexlab{a}}){Hopkins}, {Hernquist}, {Cox},
  {Di Matteo}, {Martini}, {Robertson}, \& {Springel}}]{Hopkins2005c}
{Hopkins}, P.~F., {Hernquist}, L., {Cox}, T.~J., {Di Matteo}, T., {Martini},
  P., {Robertson}, B., \& {Springel}, V. 2005{\natexlab{a}}, \apj, 630, 705

\bibitem[{{Hopkins} {et~al.}(2008{\natexlab{b}}){Hopkins}, {Hernquist}, {Cox},
  \& {Kere{\v s}}}]{hopkins2007a}
{Hopkins}, P.~F., {Hernquist}, L., {Cox}, T.~J., \& {Kere{\v s}}, D.
  2008{\natexlab{b}}, \apjs, 175, 356

\bibitem[{{Hopkins} {et~al.}(2005{\natexlab{b}}){Hopkins}, {Hernquist},
  {Martini}, {Cox}, {Robertson}, {Di Matteo}, \& {Springel}}]{hopkins2005d}
{Hopkins}, P.~F., {Hernquist}, L., {Martini}, P., {Cox}, T.~J., {Robertson},
  B., {Di Matteo}, T., \& {Springel}, V. 2005{\natexlab{b}}, \apjl, 625, L71

\bibitem[{{Hopkins} {et~al.}(2010{\natexlab{a}}){Hopkins}, {Murray},
  {Quataert}, \& {Thompson}}]{hopkins2010.maxsd}
{Hopkins}, P.~F., {Murray}, N., {Quataert}, E., \& {Thompson}, T.~A.
  2010{\natexlab{a}}, \mnras, 401, L19

\bibitem[{{Hopkins} {et~al.}(2007){Hopkins}, {Richards}, \&
  {Hernquist}}]{hopkins2007.templateqso}
{Hopkins}, P.~F., {Richards}, G.~T., \& {Hernquist}, L. 2007, \apj, 654, 731

\bibitem[{{Hopkins} {et~al.}(2010{\natexlab{b}}){Hopkins}, {Younger},
  {Hayward}, {Narayanan}, \& {Hernquist}}]{hopkins2009.ulirg}
{Hopkins}, P.~F., {Younger}, J.~D., {Hayward}, C.~C., {Narayanan}, D., \&
  {Hernquist}, L. 2010{\natexlab{b}}, \mnras, 17

\bibitem[{{Hopkins} {et~al.}(2006)}]{hopkins2006}
{Hopkins}, P.~F. {et~al.} 2006, \apjs, 163, 1

\bibitem[{{Huang} {et~al.}(2007)}]{huang2007}
{Huang}, J.~. {et~al.} 2007, \apjl, 660, L69

\bibitem[{{Hughes} {et~al.}(1998)}]{hughes1998}
{Hughes}, D.~H. {et~al.} 1998, \nat, 394, 241

\bibitem[{{Iono} {et~al.}(2006)}]{iono2006}
{Iono}, D. {et~al.} 2006, \apjl, 640, L1

\bibitem[{{Ivison} {et~al.}(2009{\natexlab{a}}){Ivison}, {Smail},
  {Papadopoulos}, {Wold}, {Richard}, {Swinbank}, {Kneib}, \&
  {Owen}}]{ivison2010}
{Ivison}, R., {Smail}, I., {Papadopoulos}, P.~P., {Wold}, I., {Richard}, J.,
  {Swinbank}, A.~M., {Kneib}, J., \& {Owen}, F.~N. 2009{\natexlab{a}}, MNRAS,
  in press [astro-ph/0912.1591]

\bibitem[{{Ivison} {et~al.}(2009{\natexlab{b}})}]{ivison2009.firradio}
{Ivison}, R.~J. {et~al.} 2009{\natexlab{b}}, \mnras, 1794

\bibitem[{{Kaviani} {et~al.}(2003){Kaviani}, {Haehnelt}, \&
  {Kauffmann}}]{kaviani2003}
{Kaviani}, A., {Haehnelt}, M.~G., \& {Kauffmann}, G. 2003, \mnras, 340, 739

\bibitem[{{Kennicutt}(1998)}]{kennicutt1998}
{Kennicutt}, Jr., R.~C. 1998, \apj, 498, 541

\bibitem[{{Kere{\v s}} {et~al.}(2009{\natexlab{a}}){Kere{\v s}}, {Katz},
  {Dav{\'e}}, {Fardal}, \& {Weinberg}}]{keres2009b}
{Kere{\v s}}, D., {Katz}, N., {Dav{\'e}}, R., {Fardal}, M., \& {Weinberg},
  D.~H. 2009{\natexlab{a}}, \mnras, 396, 2332

\bibitem[{{Kere{\v s}} {et~al.}(2009{\natexlab{b}}){Kere{\v s}}, {Katz},
  {Fardal}, {Dav{\'e}}, \& {Weinberg}}]{keres2009}
{Kere{\v s}}, D., {Katz}, N., {Fardal}, M., {Dav{\'e}}, R., \& {Weinberg},
  D.~H. 2009{\natexlab{b}}, \mnras, 395, 160

\bibitem[{{Kere{\v s}} {et~al.}(2005){Kere{\v s}}, {Katz}, {Weinberg}, \&
  {Dav{\'e}}}]{keres2005}
{Kere{\v s}}, D., {Katz}, N., {Weinberg}, D.~H., \& {Dav{\'e}}, R. 2005,
  \mnras, 363, 2

\bibitem[{{Kim} {et~al.}(2002){Kim}, {Veilleux}, \& {Sanders}}]{kim2002}
{Kim}, D., {Veilleux}, S., \& {Sanders}, D.~B. 2002, \apjs, 143, 277

\bibitem[{{Koekemoer} {et~al.}(2007)}]{koekemoer2007}
{Koekemoer}, A.~M. {et~al.} 2007, \apjs, 172, 196

\bibitem[{{Komatsu} {et~al.}(2010)}]{komatsu2010}
{Komatsu}, E. {et~al.} 2010, ApJ, in press [astro-ph/1001.4538]

\bibitem[{{Kov{\'a}cs} {et~al.}(2006)}]{kovacs2006}
{Kov{\'a}cs}, A. {et~al.} 2006, \apj, 650, 592

\bibitem[{{Kroupa}(2001)}]{kroupa2001}
{Kroupa}, P. 2001, \mnras, 322, 231

\bibitem[{{Lacki} \& {Thompson}(2009)}]{lacki2009b}
{Lacki}, B.~C. \& {Thompson}, T.~A. 2009, ApJ, submitted [astro-ph/0910.0478]

\bibitem[{{Lacki} {et~al.}(2009){Lacki}, {Thompson}, \&
  {Quataert}}]{lacki2009a}
{Lacki}, B.~C., {Thompson}, T.~A., \& {Quataert}, E. 2009, ApJ, submitted
  [astro-ph/0907.4161]

\bibitem[{{Lagache} {et~al.}(2003){Lagache}, {Dole}, \& {Puget}}]{lagache2003}
{Lagache}, G., {Dole}, H., \& {Puget}, J. 2003, \mnras, 338, 555

\bibitem[{{Le Floc'h} {et~al.}(2005)}]{lefloch2005}
{Le Floc'h}, E. {et~al.} 2005, \apj, 632, 169

\bibitem[{{Ma} {et~al.}(1998)}]{ma1998}
{Ma}, C. {et~al.} 1998, \aj, 116, 516

\bibitem[{{Magnelli} {et~al.}(2009){Magnelli}, {Elbaz}, {Chary}, {Dickinson},
  {Le Borgne}, {Frayer}, \& {Willmer}}]{magnelli2009}
{Magnelli}, B., {Elbaz}, D., {Chary}, R.~R., {Dickinson}, M., {Le Borgne}, D.,
  {Frayer}, D.~T., \& {Willmer}, C.~N.~A. 2009, \aap, 496, 57

\bibitem[{{Men{\'e}ndez-Delmestre} {et~al.}(2007)}]{menendez2007}
{Men{\'e}ndez-Delmestre}, K. {et~al.} 2007, \apjl, 655, L65

\bibitem[{{Men{\'e}ndez-Delmestre} {et~al.}(2009)}]{menendez2009}
---. 2009, \apj, 699, 667

\bibitem[{{Mihos} \& {Hernquist}(1994)}]{mihos1994}
{Mihos}, J.~C. \& {Hernquist}, L. 1994, \apjl, 431, L9

\bibitem[{{Mihos} \& {Hernquist}(1996)}]{mihos1996}
---. 1996, \apj, 464, 641

\bibitem[{{Momjian} {et~al.}(2010){Momjian}, {Wang}, {Knudsen}, {Carilli},
  {Cowie}, \& {Barger}}]{momjian2010}
{Momjian}, E., {Wang}, W., {Knudsen}, K.~K., {Carilli}, C.~L., {Cowie}, L.~L.,
  \& {Barger}, A.~J. 2010, AJ, in press [astro-ph/1002.3324]

\bibitem[{{Murphy}(2009)}]{murphy2009}
{Murphy}, E.~J. 2009, \apj, 706, 482

\bibitem[{{Murphy}(2006)}]{murphy2006}
{Murphy}, E.~J.~a. 2006, \apj, 638, 157

\bibitem[{{Murray}(2009)}]{murray2009}
{Murray}, N. 2009, \apj, 691, 946

\bibitem[{{Murray} {et~al.}(2005){Murray}, {Quataert}, \&
  {Thompson}}]{murray2005}
{Murray}, N., {Quataert}, E., \& {Thompson}, T.~A. 2005, \apj, 618, 569

\bibitem[{{Narayanan} {et~al.}(2009{\natexlab{a}}){Narayanan}, {Cox},
  {Hayward}, {Younger}, \& {Hernquist}}]{narayanan2009.co}
{Narayanan}, D., {Cox}, T.~J., {Hayward}, C.~C., {Younger}, J.~D., \&
  {Hernquist}, L. 2009{\natexlab{a}}, \mnras, 400, 1919

\bibitem[{{Narayanan} {et~al.}(2010){Narayanan}, {Hayward}, {Cox}, {Hernquist},
  {Jonsson}, {Younger}, \& {Groves}}]{narayanan2009.smg}
{Narayanan}, D., {Hayward}, C.~C., {Cox}, T.~J., {Hernquist}, L., {Jonsson},
  P., {Younger}, J.~D., \& {Groves}, B. 2010, \mnras, 401, 1613

\bibitem[{{Narayanan} {et~al.}(2006)}]{narayanan2006}
{Narayanan}, D. {et~al.} 2006, \apjl, 642, L107

\bibitem[{{Narayanan} {et~al.}(2008)}]{narayanan2008}
---. 2008, \apj, 684, 996

\bibitem[{{Narayanan} {et~al.}(2009{\natexlab{b}})}]{narayanan2009.dog}
---. 2009{\natexlab{b}}, MNRAS, submitted [astro-ph/0910.2234]

\bibitem[{{Neri} {et~al.}(2003)}]{neri2003}
{Neri}, R. {et~al.} 2003, \apjl, 597, L113

\bibitem[{{Patnaik} {et~al.}(1992){Patnaik}, {Browne}, {Wilkinson}, \&
  {Wrobel}}]{patnaik1992}
{Patnaik}, A.~R., {Browne}, I.~W.~A., {Wilkinson}, P.~N., \& {Wrobel}, J.~M.
  1992, \mnras, 254, 655

\bibitem[{{P{\'e}rez-Gonz{\'a}lez} {et~al.}(2005)}]{perezgonzalez2005}
{P{\'e}rez-Gonz{\'a}lez}, P.~G. {et~al.} 2005, \apj, 630, 82

\bibitem[{{Pope} {et~al.}(2006)}]{pope2006}
{Pope}, A. {et~al.} 2006, \mnras, 370, 1185

\bibitem[{{Pope} {et~al.}(2008)}]{pope2008b}
---. 2008, \apj, 675, 1171

\bibitem[{{Puccetti} {et~al.}(2009)}]{puccetti2009}
{Puccetti}, S. {et~al.} 2009, \apjs, 185, 586

\bibitem[{{Redman} {et~al.}(1998){Redman}, {Feldman}, \&
  {Matthews}}]{redman1998}
{Redman}, R.~O., {Feldman}, P.~A., \& {Matthews}, H.~E. 1998, \aj, 116, 1478

\bibitem[{{Robertson} {et~al.}(2004){Robertson}, {Yoshida}, {Springel}, \&
  {Hernquist}}]{robertson2004}
{Robertson}, B., {Yoshida}, N., {Springel}, V., \& {Hernquist}, L. 2004, \apj,
  606, 32

\bibitem[{{Salpeter}(1955)}]{salpeter1955}
{Salpeter}, E.~E. 1955, \apj, 121, 161

\bibitem[{{Sanders} \& {Mirabel}(1996)}]{sanders1996}
{Sanders}, D.~B. \& {Mirabel}, I.~F. 1996, \araa, 34, 749

\bibitem[{{Sanders} {et~al.}(1988){Sanders}, {Soifer}, {Elias}, {Madore},
  {Matthews}, {Neugebauer}, \& {Scoville}}]{sanders1988a}
{Sanders}, D.~B., {Soifer}, B.~T., {Elias}, J.~H., {Madore}, B.~F., {Matthews},
  K., {Neugebauer}, G., \& {Scoville}, N.~Z. 1988, \apj, 325, 74

\bibitem[{{Sanders} {et~al.}(2007)}]{sanders2007}
{Sanders}, D.~B. {et~al.} 2007, \apjs, 172, 86

\bibitem[{{Sault} {et~al.}(1995){Sault}, {Teuben}, \& {Wright}}]{sault1995}
{Sault}, R.~J., {Teuben}, P.~J., \& {Wright}, M.~C.~H. 1995, in ASP Conf. Ser.
  77: Astronomical Data Analysis Software and Systems IV, ed. R.~A. {Shaw},
  H.~E. {Payne}, \& J.~J.~E. {Hayes}, 433

\bibitem[{{Schinnerer} {et~al.}(2007)}]{schinnerer2007}
{Schinnerer}, E. {et~al.} 2007, \apjs, 172, 46

\bibitem[{{Scott} {et~al.}(2008)}]{scott2008}
{Scott}, K.~S. {et~al.} 2008, \mnras, 385, 2225

\bibitem[{{Scott} {et~al.}(2002)}]{scott2002}
{Scott}, S.~E. {et~al.} 2002, \mnras, 331, 817

\bibitem[{{Scoville} {et~al.}(2007)}]{scoville2007}
{Scoville}, N. {et~al.} 2007, \apjs, 172, 1

\bibitem[{{Scoville} {et~al.}(1991){Scoville}, {Sargent}, {Sanders}, \&
  {Soifer}}]{scoville1991}
{Scoville}, N.~Z., {Sargent}, A.~I., {Sanders}, D.~B., \& {Soifer}, B.~T. 1991,
  \apjl, 366, L5

\bibitem[{{Scoville} {et~al.}(2000)}]{scoville2000}
{Scoville}, N.~Z. {et~al.} 2000, \aj, 119, 991

\bibitem[{{Serjeant} {et~al.}(2010)}]{serjeant2010}
{Serjeant}, S. {et~al.} 2010, A\&A, in press [1002.3618]

\bibitem[{{Smail} {et~al.}(1997){Smail}, {Ivison}, \& {Blain}}]{smail1997}
{Smail}, I., {Ivison}, R.~J., \& {Blain}, A.~W. 1997, \apjl, 490, L5

\bibitem[{{Solomon} {et~al.}(1997){Solomon}, {Downes}, {Radford}, \&
  {Barrett}}]{solomon1997}
{Solomon}, P.~M., {Downes}, D., {Radford}, S.~J.~E., \& {Barrett}, J.~W. 1997,
  \apj, 478, 144

\bibitem[{{Stevens} {et~al.}(2005)}]{stevens2005}
{Stevens}, J.~A. {et~al.} 2005, \mnras, 360, 610

\bibitem[{{Swinbank} {et~al.}(2006)}]{swinbank2006}
{Swinbank}, A.~M. {et~al.} 2006, \mnras, 371, 465

\bibitem[{{Swinbank} {et~al.}(2008)}]{swinbank2008}
---. 2008, \mnras, 391, 420

\bibitem[{{Tabatabaei} {et~al.}(2007)}]{tabatabaei2007}
{Tabatabaei}, F.~S. {et~al.} 2007, \aap, 466, 509

\bibitem[{{Tacconi} {et~al.}(2006)}]{tacconi2006}
{Tacconi}, L.~J. {et~al.} 2006, \apj, 640, 228

\bibitem[{{Tacconi} {et~al.}(2008)}]{tacconi2008}
---. 2008, \apj, 680, 246

\bibitem[{{Tacconi} {et~al.}(2010)}]{tacconi2010}
---. 2010, \nat, 463, 781

\bibitem[{{Taniguchi} {et~al.}(2007)}]{taniguchi2007}
{Taniguchi}, Y. {et~al.} 2007, \apjs, 172, 9

\bibitem[{{Thompson} {et~al.}(2005){Thompson}, {Quataert}, \&
  {Murray}}]{thompson2005}
{Thompson}, T.~A., {Quataert}, E., \& {Murray}, N. 2005, \apj, 630, 167

\bibitem[{{Thompson} {et~al.}(2006){Thompson}, {Quataert}, {Waxman}, {Murray},
  \& {Martin}}]{thompson2006}
{Thompson}, T.~A., {Quataert}, E., {Waxman}, E., {Murray}, N., \& {Martin},
  C.~L. 2006, \apj, 645, 186

\bibitem[{{Valiante} {et~al.}(2007)}]{valiante2007}
{Valiante}, E. {et~al.} 2007, \apj, 660, 1060

\bibitem[{{van Dokkum}(2008)}]{VanDokkum2008}
{van Dokkum}, P.~G. 2008, \apj, 674, 29

\bibitem[{{Veilleux} {et~al.}(2002){Veilleux}, {Kim}, \&
  {Sanders}}]{veilleux2002}
{Veilleux}, S., {Kim}, D.-C., \& {Sanders}, D.~B. 2002, \apjs, 143, 315

\bibitem[{{Viero} {et~al.}(2009)}]{viero2009}
{Viero}, M.~P. {et~al.} 2009, \apj, 707, 1766

\bibitem[{{Walter} {et~al.}(2009){Walter}, {Riechers}, {Cox}, {Neri},
  {Carilli}, {Bertoldi}, {Weiss}, \& {Maiolino}}]{walter2009}
{Walter}, F., {Riechers}, D., {Cox}, P., {Neri}, R., {Carilli}, C., {Bertoldi},
  F., {Weiss}, A., \& {Maiolino}, R. 2009, \nat, 457, 699

\bibitem[{{Wang} {et~al.}(2007)}]{wang2007}
{Wang}, W.-H. {et~al.} 2007, \apjl, 670, L89

\bibitem[{{Wilson} {et~al.}(2008)}]{wilson2008}
{Wilson}, G.~W. {et~al.} 2008, \mnras, 386, 807

\bibitem[{{Younger}(2010)}]{younger2010.evnproc}
{Younger}, J.~D. 2010, in Bulletin of the American Astronomical Society,
  Vol.~41, Bulletin of the American Astronomical Society, 321--+

\bibitem[{{Younger} {et~al.}(2007)}]{younger2007}
{Younger}, J.~D. {et~al.} 2007, \apj, 671, 1531

\bibitem[{{Younger} {et~al.}(2008{\natexlab{a}})}]{younger2008}
---. 2008{\natexlab{a}}, \mnras, 387, 707

\bibitem[{{Younger} {et~al.}(2008{\natexlab{b}})}]{younger2008highres}
---. 2008{\natexlab{b}}, \apj, 688, 59

\bibitem[{{Younger} {et~al.}(2009{\natexlab{a}})}]{younger2008.egsulirgs}
---. 2009{\natexlab{a}}, \mnras, 394, 1685

\bibitem[{{Younger} {et~al.}(2009{\natexlab{b}})}]{younger2009.aztecsma}
---. 2009{\natexlab{b}}, \apj, 704, 803

\bibitem[{{Yun} {et~al.}(2001){Yun}, {Reddy}, \& {Condon}}]{yun2001}
{Yun}, M.~S., {Reddy}, N.~A., \& {Condon}, J.~J. 2001, \apj, 554, 803

\end{thebibliography}

\end{document}